# Vaccination nudges: A study of pre-booked COVID-19 vaccinations in Sweden

Carl Bonander[a], Mats Ekman[b], and Niklas Jakobsson[b] *

[a]School of Public Health and Community Medicine, University of Gothenburg, Sweden

[b]Karlstad Business School, Karlstad University, Sweden

**Abstract**

A nudge changes people's actions without removing their options or altering their incentives. During the COVID-19 vaccine rollout, the Swedish Region of Uppsala sent letters with pre-booked appointments to inhabitants aged 16–17 instead of opening up manual appointment booking. Using regional and municipal vaccination data, we document a higher vaccine uptake among 16- to 17-year-olds in Uppsala compared to untreated control regions (constructed using the synthetic control method as well as neighboring municipalities). The results highlight pre-booked appointments as a strategy for increasing vaccination rates in populations with low perceived risk.

* Niklas Jakobsson (corresponding author), Professor of Economics, Karlstad Business School, Karlstad University, 65188, Sweden. E-mail: niklas.jakobsson@kau.se. Phone: +46703939009.

# 1 Introduction

COVID-19 vaccines are offered free of charge in all rich countries, but vaccination uptake mostly falls below 80 percent (Ritchie et al., 2020). Different measures can potentially increase vaccine uptake among hesitant individuals, such as cash incentives (direct payments and lotteries) and mandatory COVID-19 certificates (Campos-Mercade et al., 2021; Mills and Rüttenauer, 2021; Barber and West, 2022). These interventions can be expensive or intrusive, and the use of interventions that alter people's behavior without changing economic incentives or regulating behavior has thus received significant interest. A common approach is the use of nudges, which change the choice architecture to steer people's choices without limiting their options (Thaler and Sunstein, 2008). Two randomized controlled trials have studied the effects of nudges on COVID-19 vaccination uptake. Dai et al. (2021) found that text-based reminders effectively increased vaccination uptake from low vaccination levels early in the vaccination rollout. In contrast, Campos-Mercade et al. (2021) found no effect of three different types of nudges in a study population with high baseline vaccination uptake (70 percent). Further, Sasaki et al. (2021) conducted an online experiment testing the effect of different messages on COVID-19 vaccination intentions, finding great importance in subtle word choices. Similar results are found in Tanaka et al. (2021). In sum, there is limited empirical evidence on the effect of nudges on COVID-19 vaccination and the conditions under which a nudge will be successful.

On July 15$^{\text{th}}$, 2021, Region Uppsala, one of Sweden's 21 regional governments and home to approximately 375,000 people, sent letters with pre-booked COVID-19 vaccination appointments to all residents aged 16 and 17. Other Swedish regions simply opened up bookings for this cohort. Similarly, other age groups could make their own appointments in Region Uppsala as well as in the rest of Sweden. The alternatives were to get vaccinated or not, and the decision by the regional authorities merely changed the default from the possibility of choosing an array of times or no time to the choice of a particular time. It remained possible not to show up (which one was not charged for) or cancel the appointment. Thus, these letters in Region Uppsala provide a real-world example of an extensive vaccination nudge.

Our aim is to study whether these pre-booked appointments increased vaccine uptake. We use two empirical strategies to identify the effect. First, we use the synthetic control method to estimate the impact in Region Uppsala compared to other (untreated) Swedish regions. Second, we estimate the impact in municipalities in Region Uppsala compared to bordering (untreated) municipalities in other regions. We find a large and statistically significant effect of Region Uppsala's nudge on vaccine uptake, regardless of which of these two methods we use.

Important features of a nudge are that it neither removes nor adds alternatives and that it does not change the utilities associated with any of the available alternatives, other than through the presentation of the options (Thaler and Sunstein, 2008). The type of nudge we study in this paper involves changing the default. Such nudges have been found to influence people's actions in a number of domains. Madrian and Shea (2001) find that the decision to participate in a pension-savings program in which the employer matches one's own contribution is made far more often when it is made the default option. Similarly, Pichert and Katsikopoulos (2008) find green energy to be a more frequently chosen option when it is the default and Li et al. (2013) report that deceased-donor organ donations increase with opt-out defaults in comparison to opt-in ones. In a recent systematic review of the literature on nudging for vaccinations, Reñosa et al. (2021) find that nudges that change defaults, giving incentives, and providing reminders have been effective in increasing flu vaccinations in some settings. In a large US field experiment Milkman et al. (2021) show that text messages on the phone increase flu vaccinations. Yokum et al. (2018) also find that mailed information letters increase flu vaccinations in the US and Chapman et al. (2016) find that pre-booked appointments for flu vaccinations increased flu vaccinations.

Pre-booked vaccination appointments could also potentially influence uptake through social effects. For instance, when teenagers know their peers have received the same letter with a pre-booked appointment, they could assume their peers are more likely to get vaccinated and therefore choose to do the same. A well-known feature of peer effects is that small changes in price (or the mental cost of making an appointment) can cause large equilibrium changes when peer consumption complements own consumption. This is because the small change will induce some portion of the peer group to get vaccinated, which, through the complement, causes some other portion to get vaccinated, etc. (Becker and Murphy, 2000; see Sasaki et al., 2022, for other ways in which social effects can effectuate nudges).

Our results are consistent with those of Löfgren and Nordblom (2020), who construct a theoretical model that predicts the circumstances in which a nudge is likely to be effective. Their model shows that the likelihood of a nudge having an effect is higher for choices that the individual believes are unimportant. For choices that the individual considers important, nudges are less likely to have an effect. Since 16–17-year-olds are unlikely to suffer or die from either a COVID-19 infection (Kolk et al., 2021) or the side-effects of a vaccine (Patone et al., 2021), we should expect a larger effect from a nudge in this age group than among older individuals. Indeed, the effect that we find is greater than those based on modest monetary payments or conditional cash lotteries in previous research. We note, however, that the young age of the individuals in our study also makes it unclear how generalizable our findings are to older individuals, for whom the incentive to get vaccinated is greater.

Apart from adding to the health-economics literature on incentives and vaccination uptake (Campos-Mercade et al., 2021; Dai et al., 2021; Mills and Rüttenauer, 2021; Barber and West, 2022), we also contribute to the burgeoning literature on nudging (Madrian and Shea, 2001; Pichert and Katsikopoulos, 2008; Li et al., 2013; Löfgren and Nordblom 2020).

## 2 Methods and data
### 2.1 Empirical framework

Since we have access to both regional and municipal vaccination data, we conduct analyses on both levels to assess the effect of the nudging intervention in region Uppsala. In the regional analyses, we measure the impact through a comparative case-study approach that compares the trend in vaccination uptake between Region Uppsala and a set of untreated but similar regions. Specifically, we implement the synthetic control method to construct a synthetic Uppsala, which closely resembles the real Uppsala in terms of pre-intervention characteristics, from a combination of all other Swedish regions (Abadie et al., 2010; Abadie, 2021). The synthetic control method is designed to estimate the impact of policy interventions affecting one unit (e.g., country, region, or municipality) when only a small number of control units are available. It is a data-driven approach for estimating counterfactuals—i.e., what would have happened without the nudge—which automatically determines the weighted combination of untreated regions that provides the best match to the treated region with regard to pre-intervention outcomes and covariates. The weighted average vaccination uptake from the synthetic control group then provides the counterfactual trend of the vaccination share for Region Uppsala; i.e., it predicts how the vaccination rates would have turned out in the absence of the nudging intervention. For a detailed presentation of the method, see Abadie (2021). For recent implementations of the synthetic control method related to the COVID-19 pandemic, see, for example, Cho (2020), Mitze (2020), and Alfano et al. (2021).

Abadie and Gardeazabal (2003) propose a nested optimization routine to simultaneously determine (i) a set of unit weights (one for each control) that determine each untreated unit's contribution to the synthetic control and (ii) variable importance weights (one for each covariate) to prioritize a good match on strong predictor outcomes. The latter aspect is useful in small datasets where a perfect match cannot be expected for all included variables. We also consider equal importance weights in sensitivity analyses. Following Abadie et al. (2010), we make inference using in-place placebo studies, where we estimate "effects" in each control region to assess uncertainty. Thus, we can assess if the effect in the treated region is large relative to the estimated effects in the non-treated regions.

In the municipal analysis, we compare the eight municipalities in Region Uppsala to all eight municipalities that share a border with a treated municipality. The geographic proximity between municipalities just within and just outside Region Uppsala makes comparisons between municipalities fruitful for detecting differences in vaccination uptake due to the intervention by Region Uppsala. Four of the eight municipalities of Region Uppsala (Uppsala, Tierp, Östhammar, and Älvkarleby) border only one municipality outside of the region, while the others border two or, in one case (Heby), four. The idea is that the geographical proximity should make the untreated neighbors a reasonable control group, as individuals on different sides of the border share similar social environments. Variants of this kind of identification strategy have been used extensively in studying effects of local policy variations (e.g. Card and Krueger 1994, Boone et al. 2021).

We conduct a descriptive comparison of the vaccination development in the treated and neighboring municipalities and compare the final vaccination share in the treated municipalities to those of their neighbors. We also run ordinary least squares regressions with the share of vaccinated individuals as the outcome variable, with neighbor fixed effects and covariates (see next section for details). Finally, we perform difference-in-differences and event-study difference-in-differences estimation (Schmidheiny and Siegloch 2019), in which we contrast the increase in vaccinations in the treated municipalities with the increase in the neighboring municipalities.

## 2.2 Data

The outcome data in the present study are the share of vaccinated individuals, obtained from the Public Health Agency of Sweden and structured as regional-level weekly panel data covering all 21 Swedish regions (defined in Eurostat's Nomenclature of Territorial Units for Statistics [NUTS3]). Our data contain the share of vaccinated individuals in the 16–17-year age group and the 18–19-year age group from week 1 to 46 in 2021 for all 21 Swedish regions.

We include several covariates that may be important confounders following recent empirical findings on relevant predictors of adolescents' attitudes towards COVID-19 vaccination (Fazel et al. 2021; Nivette et al. 2021; ONS 2021): the share of COVID-19 deaths in 2020 (from the Public Health Agency of Sweden); the share of the population with at least three years of higher education in 2020 (Statistics Sweden); the share of foreign-born individuals living in the region in 2020 (Statistics Sweden); the share of population that has received financial aid at any point during the past year in 2020 (Kolada, www.kolada.se); number of adolescents 10-24 per 100000 inhabitants who have received care due to alcohol addiction in 2019 (Kolada); the share of the population that has access to a fast broadband connection (100 Mbit/s) in 2020 (Kolada); the share of the population with high trust in how the Swedish health care system handled the COVID-19 pandemic in 2020 (Kolada); and NEETs (the share of 17-24 year olds that neither study nor work) in 2020 (Kolada). We also include the pre-intervention share of vaccinated 18–29-year-olds as a proxy for the general willingness to get vaccinated in each region. All variables refer to the entire population unless otherwise noted.

For the vaccination data, cells with three or fewer observations were set to zero by the Public Health Agency of Sweden due to integrity reasons, so we have some missing data in weeks when very few individuals got vaccinated. This was more of a problem in the early stage of the pandemic, when mainly individuals with medical risks (e.g., chronic lung disease, cancer, and diabetes) in the 16–17 year age group were vaccinated.

For the municipal comparison of neighboring municipalities, we use data on the vaccination share in two-week intervals and the total vaccination share in week 49. We received two-week (instead of one-week) data to reduce the number of cells with three or fewer observations. We also use the cumulative share of vaccinated individuals in week 49; with this outcome, we lose no data but do not have the time series. Municipal-level data were available for the following control variables: the share of foreign-born, the share with at least three years in higher education (both from Statistics Sweden), and the share of COVID-19 deaths in 2020 reported by the Public Health Agency of Sweden.

# 3 Results
## 3.1 Regional analyses
Figure 1 plots the trends in the share of first-dose vaccinations among 16–17-year-olds in Uppsala and the rest of Sweden. The vertical line indicates when Region Uppsala sent out letters with pre-booked vaccination times to all 16–17-year-olds (week 28). In the final week that we observe (week 46), we can see that vaccinations reached 85 percent of the age group in Uppsala and 75 percent in the other regions (unweighted average). In Table 1, we can see that Uppsala clearly differs from the average of the 20 control regions in terms of pre-intervention characteristics. It has a larger foreign-born population, higher education level, fewer NEETs, and less trouble with adolescent alcohol addiction. Synthetic Uppsala more closely matches real Uppsala on predictors with high variable importance weights. Table 2 displays the region weights for synthetic Uppsala, which are a weighted combination of four regions: Östergötland, Kronoberg, Stockholm, and Västerbotten.

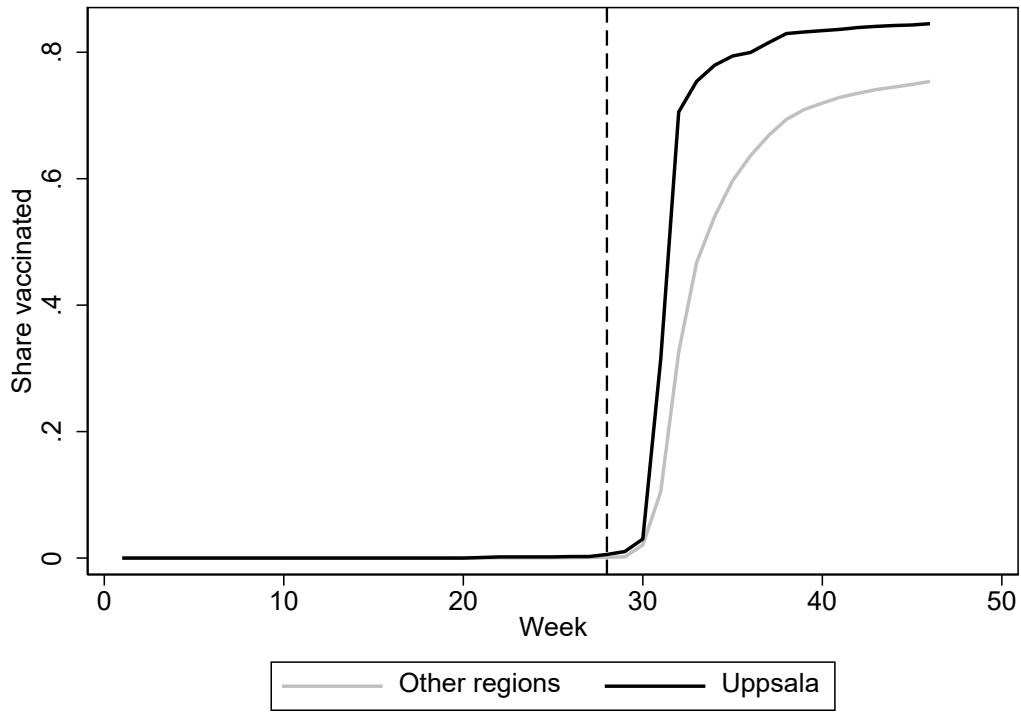

**Figure 1.** First-dose vaccinations in Uppsala (treated) and average of all other 20 Swedish regions

Table 1. Vaccination share predictor means

|  | Uppsala | Synthetic Uppsala | Average of 20 Control regions | V |
|---|---|---|---|---|
| Share foreign-born | .189 | .182 | .160 | .361 |
| Share high education | .180 | .157 | .138 | .033 |
| Share of COVID-19 deaths | .001 | .001 | .001 | .010 |
| Share with financial aid | .041 | .042 | .040 | .090 |
| Alcohol addiction per 100k | 54.0 | 69.6 | 89.1 | .044 |
| Share with fast internet | .862 | .868 | .836 | .265 |
| Share with high trust | .752 | .750 | .737 | .076 |
| Share NEETs | .062 | .070 | .078 | .085 |
| Share vaccinated (18–29 y) | .077 | .075 | .074 | .035 |

*Notes:* The period for each predictor is 2020, except for Share vaccinated (18–29 y), which refers to the mean share for all pre-intervention weeks. Variable importance weights (V) were determined via the standard synthetic control procedure.

Table 2. Region weights in synthetic Uppsala

| Region | Weight | Region | Weight |
|---|---|---|---|
| Stockholm | .133 | Västra Götaland | 0 |
| Södermanland | 0 | Värmland | 0 |
| Östergötland | .539 | Örebro | 0 |
| Jönköping | 0 | Västmanland | 0 |
| Kronoberg | .224 | Dalarna | 0 |
| Kalmar | 0 | Gävleborg | 0 |
| Gotland | 0 | Västernorrland | 0 |
| Blekinge | 0 | Jämtland | 0 |
| Skåne | 0 | Västerbotten | .104 |
| Halland | 0 | Norrbotten | 0 |

The left panel in Figure 2 shows the difference in the share of first-dose vaccinations between Uppsala and synthetic Uppsala for the treated age group (16–17 years old). There is a clear difference in the share vaccinated between Uppsala and synthetic Uppsala in the post-treatment period, which peaks in week 32 at 34.3 percentage points. In week 46, the final week of measurements, the difference is 11.7 percentage points (72.8 in synthetic Uppsala and 84.5 in actual Uppsala). In the middle panel in Figure 2, we compare the effect estimated for Uppsala with the effect of placebo interventions implemented in the other 20 regions. Reassuringly, we can see that no other region has an effect estimate close to the one in Uppsala. The right panel in Figure 2 ranks the post-intervention effect sizes across all regions, showing that Uppsala has by far the largest estimated effect (with a placebo-based p-value of 1/21=0.048). Abadie et al. (2010) suggest using the ratio between the post-to-pre-intervention root mean squared error (RMSE) in the outcome variable (vaccination uptake) to handle differences in pre-intervention fit across the placebo analyses when assessing significance, which is neither feasible nor necessary with our data given that the RMSE in the pre-intervention period is zero in almost all analyses. Overall, the analysis implies a large and persistent effect of the intervention.

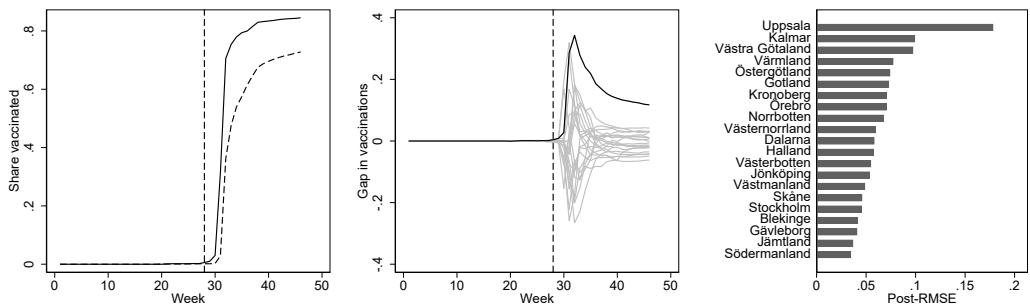

**Figure 2.** Effect (left), placebo (middle), and post-intervention effect size (right) plots. The left panel shows the share of first-dose vaccination by week in Uppsala (black) and synthetic Uppsala (dashed) among 16–17-year-olds. The middle panel shows effects estimated by assessing the vaccination share gaps between Uppsala and its synthetic counterpart (black) and equivalently-defined placebo gaps in all 20 control regions (gray). The right panel shows the post-intervention root mean squared error (RMSE) in vaccination uptake from the synthetic control analysis in Uppsala and all other regions.

To further scrutinize the findings, we assess the vaccination share for the age group 18–29 years in the Uppsala region. Since they were not treated with pre-booked appointments, we do not expect them to have a higher vaccination rate than the same age group in synthetic Uppsala. However, there may be spillovers in the treatment; increased vaccinations in the treated age group may increase vaccinations among friends and relatives in the older age group. Appendix Figure A1 shows the results for the age group 18–29 years (predictor means and weights in Appendix Tables A1 and A2). In the final week, the difference is 4.6 percentage points (80.3 in synthetic Uppsala and 84.9 in actual Uppsala). With a placebo-based p-value of 0.524, we interpret this difference as a chance finding. In Appendix Figure A2 (predictor means and weights in Appendix Tables A3 and A4), we estimate the effect on second-dose vaccinations for 16–17-year-olds, which yields similar results as in our main analysis (placebo-based p-value: 0.048).

A potential concern with our application of the synthetic control method is that there is limited variation in the pre-intervention outcomes. Since the variable importance algorithm proposed by Abadie & Gardazeabal (2003) uses pre-intervention outcomes to estimate variable importance weights, the variable weights in the main model may be inconsistent. As a sensitivity analysis, we therefore re-run the optimization while assigning equal importance to all predictors. The results, which are presented in Appendix Figure A3 and Appendix Tables A5-A6, are reassuringly very similar to the main specification. Another concern is that the results may be sensitive to our choice of included covariates. Appendix Figure A4 shows results from analyses where we successively excluded the confounder with the highest variable weight in Table 1, resulting in nine specifications including one to nine confounding variables. All estimates (gray lines) closely track each other, and the main finding therefore seems robust to the choice of included covariates. Synthetic control analyses may also be sensitive to how one chooses to input the pre-intervention outcomes in the optimization (e.g., a raw mean, every time point, every other time point, etc). Following Ferman et al. (2020), we estimate 14 different specifications where we vary how we input the pre-intervention outcomes in the synthetic control optimization and find that the effect size is largest in Uppsala in all 14 specifications (Appendix Table A7). In Figure A5, we report the results of a leave-one-region-out robustness analysis, as suggested by Abadie (2021). One at a time, we take out each of the four regions that contribute to the synthetic control in Table 2 to ensure that the results are not entirely driven by the inclusion of a specific control region. All leave-one-out estimates (gray lines) closely track the findings from the main analysis. Thus, the main finding appears robust.

### 3.2 Municipal analyses

Figure 3 shows vaccination shares for 16–17-year-olds in the treated municipalities (Enköping, Heby, Håbo, Knivsta, Tierp, Uppsala, Älvkarleby, and Östhammar) and untreated neighboring municipalities (Avesta, Gävle, Norrtälje, Sala, Sandviken, Sigtuna, Upplands-Bro, and Västerås) for the final available week of data (week 49). The share of vaccinated 16-17-year-olds was 85.1 (95% confidence interval [CI]: 83.2, 87.0) percent in the treated municipalities compared with 72.2 (95% CI: 68.3, 76.1) percent in the neighboring untreated municipalities, a difference of 12.9 percentage points (95% CI: 9.0, 16.8).

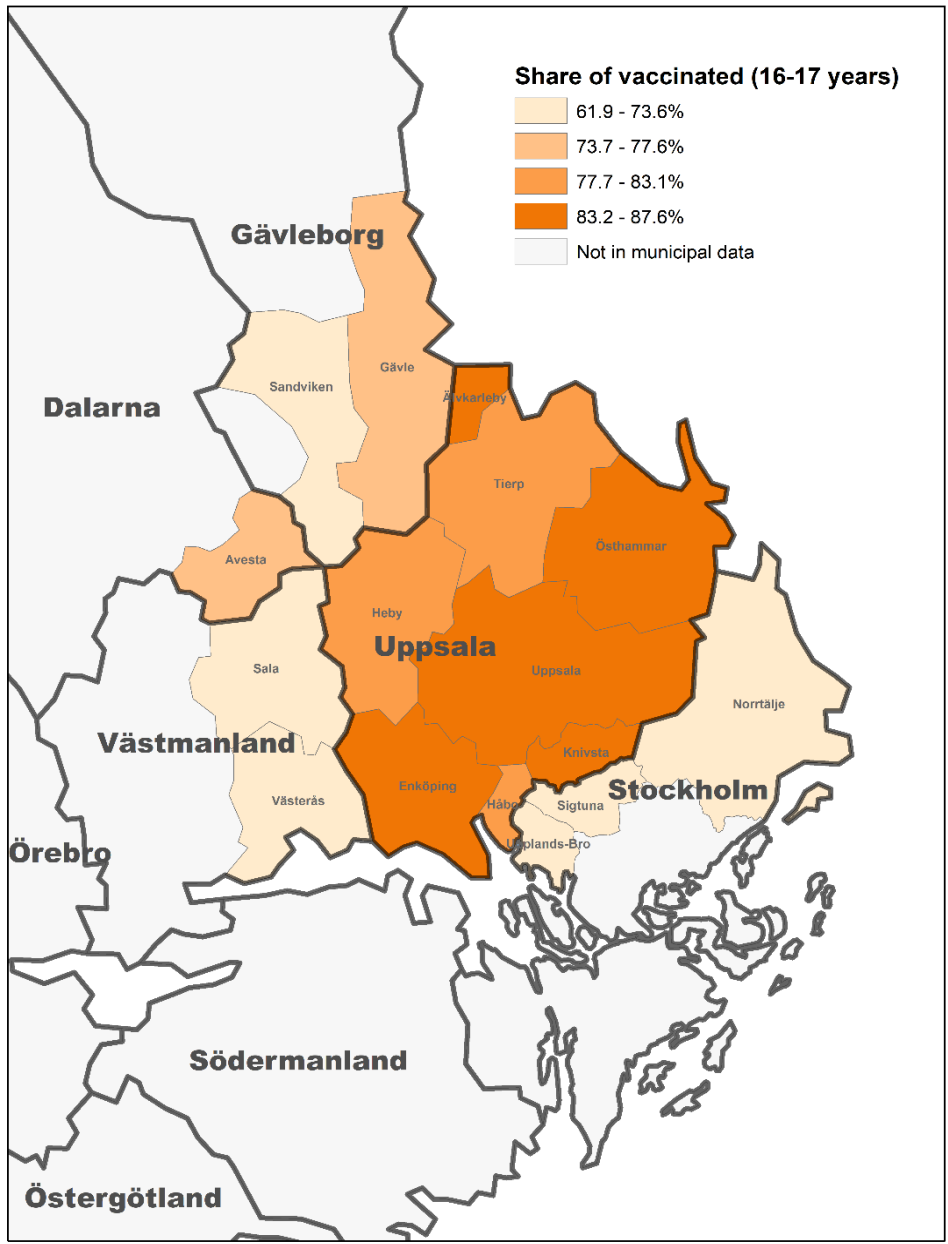

**Figure 3.** Share of vaccinated 16-17-year-olds in the treated and neighboring municipalities

Figure 4 plots the trends in the share of first-dose vaccinations among 16–17-year-olds in the treated municipalities and the neighboring municipalities.

The vertical line indicates when Region Uppsala sent out letters with pre-booked vaccination times to all 16–17-year-olds (week 28). As in the regional analysis, the share of vaccinated individuals is considerably higher in the treated municipalities compared to their untreated neighbors.

Appendix Table A8 compares summary statistics for the observed covariates between each treated municipality and their neighbors. We find meaningful differences in some neighbor groups, indicating that it is important to adjust the municipal comparisons for observables even within neighbor groups. In Table 3, we present the results from ordinary least squares (OLS) regressions, with the cumulative vaccination share in week 49 among 16–17-year-olds as the dependent variable. Because our data are spatially structured the table also includes a check for spatial autocorrelation (Moran's I) and contains both OLS and Conley corrected standard errors (Colella et al., 2919; Conley, 1999; Moran, 1950). In column 1, we include only a treatment dummy; in column 2, we include three control variables (share foreign-born, share high education, and COVID-19 deaths); in column 3, we include neighbor indicators (a dummy variable for each treated municipality, indicating which municipalities from neighboring regions it shares a border with); and in column 4, we include all of the above. The treatment estimate is not statistically different from zero in column 4 (p=0.11), but the point estimate is still considerable, and we must consider the limited degrees of freedom in a model with 16 observations and 12 control variables. The estimated treatment effect varies from 7.3 to 12.9 percentage points. As in the regional analysis, we find no evidence of an effect on vaccine uptake among 18-29-year-olds after adjusting for observable confounders (Appendix Table A9). This result suggests that the residual confounding within neighbor groups is small after adjusting for observables, assuming the same sources of bias are present in both age groups (see e.g. Lipsitch et al., 2010).

A difference-in-differences estimation (without control variables) suggests an average effect in the post-treatment period of 15.6 percentage points (95% CI: 11.6, 20.0; CI computed using wild cluster bootstrap (Cameron et al., 2008)). Figure 5 contains time-specific effect estimates and 95 percent CIs (i.e., an event study difference-in-differences estimation), showing how the effect changes over time according to the municipality-level data. Like the regional analysis, we can see that the treatment effect is massive early on, and although it decreases over time, the vaccination share in the treated municipalities is considerably higher in the final time period.

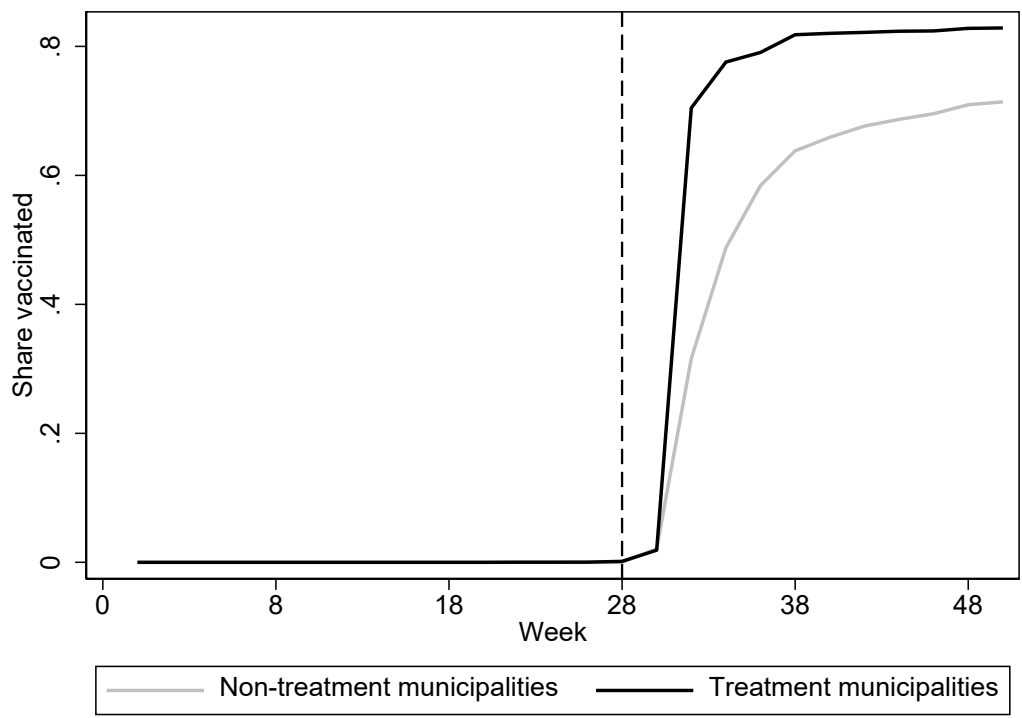

**Figure 4.** First-dose vaccinations in treated municipalities (located in Uppsala) and in their neighboring municipalities (outside Uppsala).

Table 3. Determinants of share of vaccinated 16-17-year-olds in treated and neighboring municipalities

|  | (1) | (2) | (3) | (4) |
|---|---|---|---|---|
| Treatment | 0.129*** | 0.094*** | 0.109*** | 0.073 |
|  | (0.018) | (0.014) | (0.026) | (0.028) |
| Neighbor indicators | No | No | Yes | Yes |
| Share foreign-born | No | -0.531*** | No | -0.418 |
|  |  | (0.114) |  | (0.273) |
| Share high education | No | 0.215** | No | 0.150 |
|  |  | (0.084) |  | (0.181) |
| COVID-19 deaths | No | 6.375 | No | -4.999 |
|  |  | (5.195) |  | (17.053) |
| Constant | 0.722*** | 0.774*** | 0.761*** | 0.843*** |
|  | (0.013) | (0.027) | (0.034) | (0.087) |
| $R^2$ | 0.782 | 0.900 | 0.934 | 0.976 |
| Moran's I (residuals) | 0.093 | -0.191 | -0.198 | -0.409** |
| Conley SE (treatment) | 0.019*** | 0.010*** | 0.013*** | 0.013*** |

Notes: The dependent variable is the share of 16–17-year-olds vaccinated in week 49 in the 16 included municipalities. Ordinary least squares regressions controlling for Treatment (pre-booked appointments), Neighbor indicators (one dummy variable for each treated municipality, indicating its neighbors), as well as the control variables Share foreign-born, Share high education, and COVID-19 deaths. Moran's I for spatial residual autocorrelation and Conley standard errors accounting for spatial autocorrelation were computed assuming a maximum distance for spatial autocorrelation of 65 kilometers (the minimum distance from which all area centroids shared at least one neighbor) and a Bartlett kernel using the *acreg* package for Stata. Other distance choices and a uniform kernel led to similar results, but standard errors could not be computed in Model 4 using a uniform kernel. * $p<0.1$, ** $p<0.05$, *** $p<0.01$.

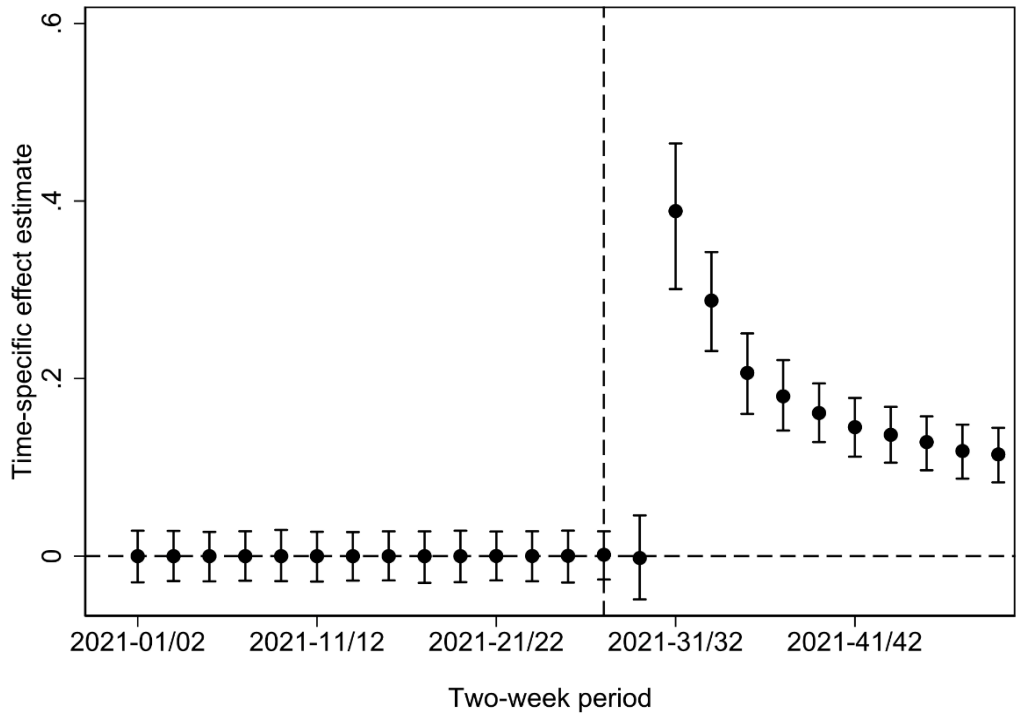

**Figure 5.** Time-specific coefficients and 95% wild cluster bootstrap confidence intervals from the difference-in-differences estimation

## 4 Discussion and conclusion

Our regional analysis suggests that pre-booked vaccination appointments increased vaccination uptake among 16–17-year-olds in the Uppsala region by about 11.7 percentage points, compared to a counterfactual uptake of 72.8 percent (in week 46). The municipal analyses also suggest an effect of 7.3–12.9 percentage points (in week 49). Although our estimates may be biased due to unobserved confounding, they are substantial, robust, and specific to the treated age range over two identification methods and datasets. They are also theoretically plausible. Our effect estimate is considerably larger than the effects found for modest monetary payments or conditional cash lotteries to increase COVID-19 vaccinations. Campos-Mercade et al. (2021) find that a monetary payment of 200 Swedish kronor (about $24) increased vaccinations by 4.2 percentage points (from a baseline of 71.6 percent) in a random sample of Swedes aged 18–49 years. Barber and West (2021) report that a conditional cash lottery in Ohio increased the vaccination share in the state population by 0.7 percentage points. In a study on nudges to increase COVID-19 vaccination uptake, Dai et al. (2021) find that text-based reminders can effectively increase vaccination uptake from low initial vaccination levels in the overall population, at least in the early stages

of the vaccination rollout. Conversely, Campos-Mercade et al. (2021) find no effects of three different nudges on COVID-19 vaccination uptake when vaccination uptake is already above 70 percent.

It may be that the effect is more pronounced in younger age groups. Löfgren and Nordblom (2020) argue that nudges should be more effective for choices that are considered unimportant by the individuals making them. Since 16–17-year-olds are unlikely to suffer from severe illness or death in case of a COVID-19 infection, whereas the risk is considerably higher for older individuals (Kolk et al., 2021), we should not expect the effect of the pre-booked vaccination appointments to be as large in the general population. Additionally, while previous studies consider lighter nudges, we study the impact of a nudge that changes the default alternative, something that has been shown to be impactful when considering choices in other domains (e.g., Madrian and Shea, 2001; Pichert and Katsikopoulos, 2008; Li et al., 2013). However, one should remember that this is a case study and the findings for Uppsala may not be generalizable to other Swedish regions or to regions in other countries.

In summary, pre-booked appointments seem to provide a simple and effective nudge that could be used more broadly to increase vaccine uptake in the future (e.g., for COVID-19 booster doses or vaccinations for other viruses).

## Acknowledgments
We would like to thank seminar participants at Örebro University, Jönköping University, the EuHEA Conference, and Siri Jakobsson Støre for providing useful comments and suggestions. Mats Ekman and Niklas Jakobsson acknowledges funding from Jan Wallanders and Tom Hedelius stiftelse & Tore Browaldhs stiftelse (grant number P22-0018). Carl Bonander acknowledges funding from the Swedish Research Council for Health, Working-Life, and Welfare (Forte; grant number 2020-00962).

## Declaration of competing interests
The authors declare that they have no competing financial interests or personal relationships that have influenced the work reported in this article.

## Credit author statement
Carl Bonander: Methodology, Validation, Formal Analysis, Data Curation, Writing – Review & Editing, Visualization. Mats Ekman: Conceptualization,

Methodology, Data Curation, Writing – Review & Editing. Niklas Jakobsson: Conceptualization, Methodology, Validation, Formal Analysis, Data Curation, Writing – Original Draft, Writing – Review & Editing, Project Administration.

# Appendix

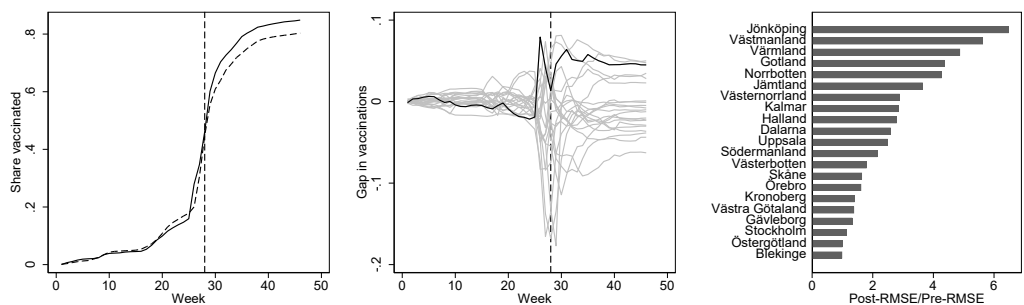

**Figure A1.** Effect (left), placebo (middle) and post-intervention effect size (right) plots for first-dose vaccinations among 18-29-year-olds (the untreated age group). The left panel shows the share of first-dose vaccinations by week in Uppsala (black) and synthetic Uppsala (dashed). The middle panel shows effects estimated by assessing the vaccination share gaps between Uppsala and its synthetic counterpart (black) and equivalently defined placebo gaps in all 20 control regions (gray). The right panel shows the ratio between post-intervention root mean squared error (RMSE) to the pre-intervention RMSE from the synthetic control analysis in Uppsala and all other regions. The specification in this analysis differs from our main analysis in that we standardize the effect sizes in the right panel by the pre-intervention RMSE to account for the fact that the pre-intervention fit can vary across regions, as suggested by Abadie et al. (2010). This was not possible nor necessary in our main analysis due to the limited variation in the outcome before vaccinations were introduced among 16-17-year-olds.

**Table A1.** Vaccination share predictor means

|  | Uppsala | Synthetic Uppsala | Average of 20 Control regions | V |
|---|---|---|---|---|
| Share foreign-born | .189 | .170 | .160 | .028 |
| Share high education | .180 | .162 | .138 | .031 |
| Share of COVID-19 deaths | .001 | .001 | .001 | .037 |
| Share with financial aid | .041 | .040 | .040 | .097 |
| Alcohol addiction per 100k | 54.0 | 65.1 | 89.1 | .077 |
| Share with fast internet | .862 | .883 | .836 | .024 |
| Share with high trust | .752 | .765 | .737 | .076 |
| Share NEETs | .062 | .069 | .078 | .137 |
| Share vaccinated (18–29 y) | .077 | .077 | .074 | .514 |

*Notes:* The period for each predictor is 2020, except for Share vaccinated (18–29 y), which refers to the mean share for all pre-intervention weeks. Variable importance weights (V) were determined via the standard synthetic control procedure.

**Table A2.** Region weights in synthetic Uppsala

| Region | Weight | Region | Weight |
|---|---|---|---|
| Stockholm | .136 | Västra Götaland | 0 |
| Södermanland | 0 | Värmland | 0 |
| Östergötland | .697 | Örebro | 0 |
| Jönköping | 0 | Västmanland | 0 |
| Kronoberg | .224 | Dalarna | 0 |
| Kalmar | 0 | Gävleborg | 0 |
| Gotland | 0 | Västernorrland | 0 |
| Blekinge | 0 | Jämtland | 0 |
| Skåne | 0 | Västerbotten | .167 |
| Halland | 0 | Norrbotten | 0 |

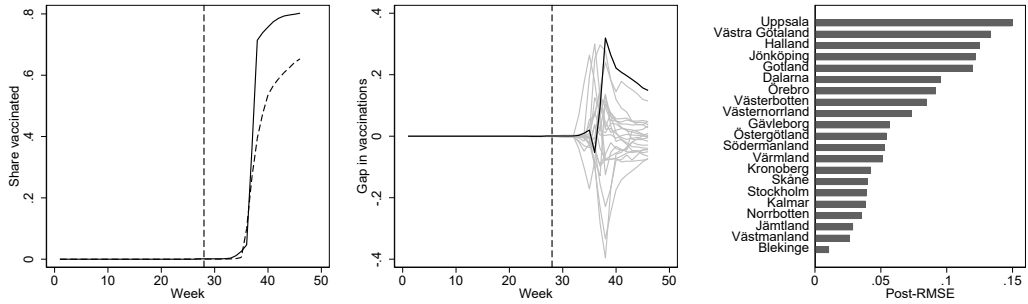

**Figure A2.** Effect (left), placebo (middle) and post-intervention effect size (right) plots for second-dose vaccinations among 16-17-year-olds. The left panel shows the share of second-dose vaccinations by week in Uppsala (black) and synthetic Uppsala (dashed). The middle panel shows effects estimated by assessing the vaccination share gaps between Uppsala and its synthetic counterpart (black) and equivalently defined placebo gaps in all 20 control regions (gray). The right panel shows the post-intervention root mean squared error (RMSE) from the synthetic control analysis in Uppsala and all other regions.

Table A3. Vaccination share predictor means

|  | Uppsala | Synthetic Uppsala | Average of 20 Control regions | V |
|---|---|---|---|---|
| Share foreign-born | .189 | .177 | .160 | .339 |
| Share high education | .180 | .162 | .138 | .065 |
| Share of COVID-19 deaths | .001 | .001 | .001 | .013 |
| Share with financial aid | .041 | .037 | .040 | .097 |
| Alcohol addiction per 100k | 54.0 | 86.4 | 89.1 | .031 |
| Share with fast internet | .862 | .873 | .836 | .024 |
| Share with high trust | .752 | .756 | .737 | .001 |
| Share NEETs | .062 | .069 | .078 | .237 |
| Share vaccinated (18–29 y), dose two | .029 | .030 | .030 | .026 |

*Notes:* The period for each predictor is 2020, except for Share vaccinated (18–29 y), which refers to the mean share for all pre-intervention weeks. Variable importance weights (V) were determined via the standard synthetic control procedure.

Table A4. Region weights in synthetic Uppsala

| Region | Weight | Region | Weight |
|---|---|---|---|
| Stockholm | .179 | Västra Götaland | 0 |
| Södermanland | 0 | Värmland | 0 |
| Östergötland | .374 | Örebro | 0 |
| Jönköping | 0 | Västmanland | 0 |
| Kronoberg | .193 | Dalarna | 0 |
| Kalmar | 0 | Gävleborg | 0 |
| Gotland | 0 | Västernorrland | 0 |
| Blekinge | 0 | Jämtland | 0 |
| Skåne | 0 | Västerbotten | .254 |
| Halland | 0 | Norrbotten | 0 |

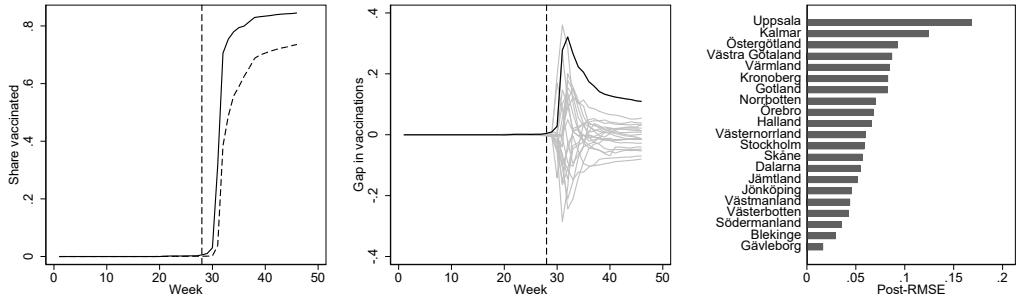

**Figure A3.** Effect (left), placebo (middle) and post-intervention effect size (right) plots, using equal weights for the included covariates. The left panel shows the share of first-dose vaccination by week in Uppsala (black) and synthetic Uppsala (dashed) among 16-17-year-olds. The middle panel shows effects estimated by assessing the vaccination share gaps between Uppsala and its synthetic counterpart (black) and equivalently defined placebo gaps in all 20 control regions (gray). The right panel shows the post-intervention root mean squared error (RMSE) from the synthetic control analysis in Uppsala and all other regions.

**Table A5.** Vaccination share predictor means

|  | Uppsala | Synthetic Uppsala | Average of 20 Control regions | V |
|---|---|---|---|---|
| Share foreign-born | .189 | .179 | .160 | .111 |
| Share high education | .180 | .164 | .138 | .111 |
| Share of COVID-19 deaths | .001 | .001 | .001 | .111 |
| Share with financial aid | .041 | .040 | .040 | .111 |
| Alcohol addiction per 100k | 54.0 | 62.7 | 89.1 | .111 |
| Share with fast internet | .862 | .887 | .836 | .111 |
| Share with high trust | .752 | .757 | .737 | .111 |
| Share NEETs | .062 | .070 | .078 | .111 |
| Share vaccinated (18–29 y) | .077 | .075 | .074 | .111 |

*Notes:* The period for each predictor is 2020, except for Share vaccinated (18–29 y), which refers to the mean share for all pre-intervention weeks. Variable importance weights (V) were determined via the standard synthetic control procedure.

**Table A6.** Region weights in synthetic Uppsala

| Region | Weight | Region | Weight |
|---|---|---|---|
| Stockholm | .177 | Västra Götaland | 0 |
| Södermanland | 0 | Värmland | 0 |
| Östergötland | .74 | Örebro | 0 |
| Jönköping | 0 | Västmanland | 0 |
| Kronoberg | 0 | Dalarna | 0 |
| Kalmar | 0 | Gävleborg | 0 |
| Gotland | 0 | Västernorrland | 0 |
| Blekinge | 0 | Jämtland | 0 |
| Skåne | 0 | Västerbotten | .083 |
| Halland | 0 | Norrbotten | 0 |

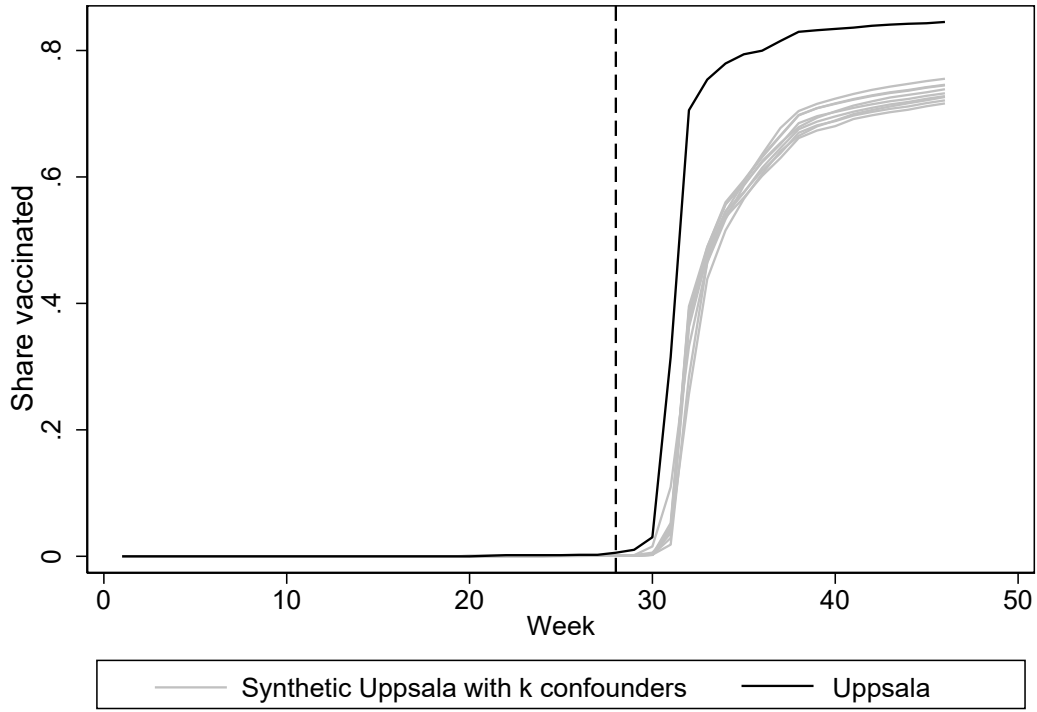

**Figure A4.** Leave-k-confounders-out estimates of first-dose vaccination

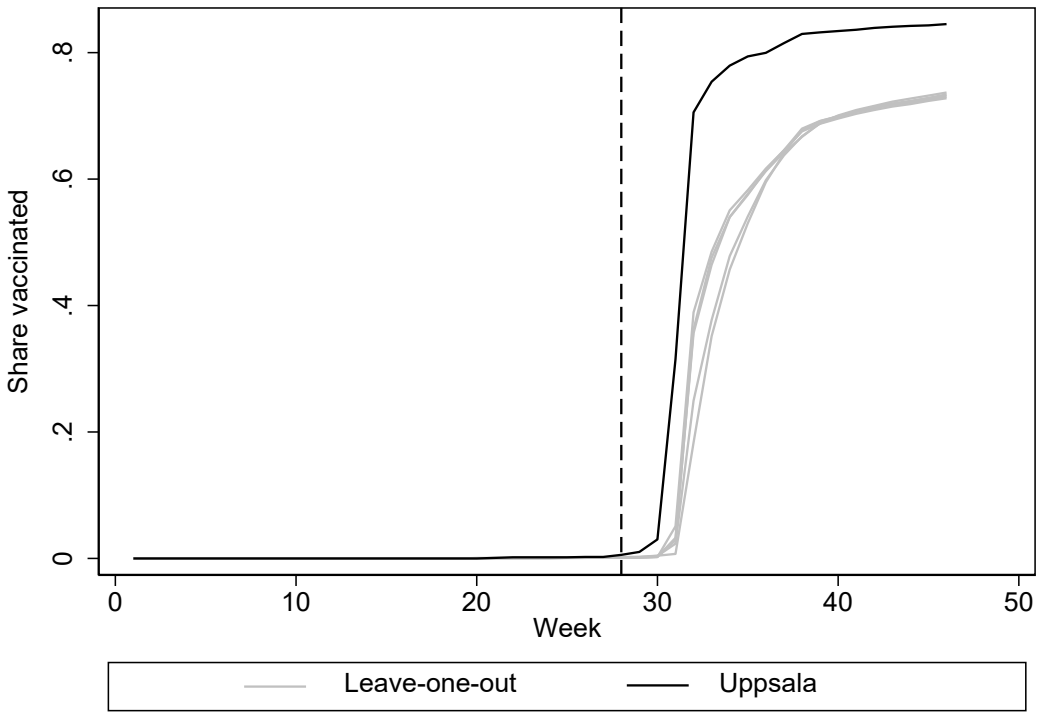

**Figure A5.** Leave-one-region-out estimates of first-dose vaccination

**Table A7**. Specifications searching

| Specification | (1a) | (1b) | (2a) | (2b) | (3a) | (3b) | (4a) |
|---|---|---|---|---|---|---|---|
| p-value | 0.048 | 0.048 | 0.048 | 0.048 | 0.048 | 0.048 | 0.048 |
| Specification | (4b) | (5a) | (5b) | (6a) | (6b) | (7a) | (7b) |
| p-value | 0.048 | 0.048 | 0.048 | 0.048 | 0.048 | 0.048 | 0.048 |

Notes: Since our data have (almost) no variation in the outcome variable in any region before the vaccination rollout, we use the vaccination share among 18-29 year olds as the main variable to match on. Specifications refer to: (1) all pre-treatment vaccination shares among 18-29 year olds, (2) the first three-fourths of the values, (3) the first half of the values, (4) odd pre-treatment values, (5) even pre-treatment values, (6) pre-treatment mean, and (7) three values. Specifications ending with b includes all additional eight covariates, while specifications ending with a, includes no additional covariates. The post-intervention effect size is largest in Uppsala in each specification, p-value=1/21=0.048.

**Table A8.** Control variable comparison between treated municipalities and their neighboring municipalities

|  | Treated municipality | Average of neighboring municipalities |
|---|---|---|
|  | Enköping | Sala, Västerås |
| Share foreign-born | 0.161 | 0.189 |
| Share high education | 0.208 | 0.230 |
| Share of Covid-19 deaths | 0.00256 | 0.00272 |
|  | Håbo | Sigtuna, Upplands-Bro |
| Share foreign-born | 0.157 | 0.326 |
| Share high education | 0.193 | 0.218 |
| Share of Covid-19 deaths | 0.00107 | 0.00270 |
|  | Knivsta | Norrtälje, Sigtuna |
| Share foreign-born | 0.144 | 0.245 |
| Share high education | 0.369 | 0.178 |
| Share of Covid-19 deaths | 0.00112 | 0.00263 |
|  | Uppsala | Norrtälje |
| Share foreign-born | 0.221 | 0.135 |
| Share high education | 0.423 | 0.168 |
| Share of Covid-19 deaths | 0.00187 | 0.00254 |
|  | Östhammar | Norrtälje |
| Share foreign-born | 0.096 | 0.135 |
| Share high education | 0.149 | 0.168 |
| Share of Covid-19 deaths | 0.00230 | 0.00254 |
|  | Tierp | Gävle |
| Share foreign-born | 0.131 | 0.159 |
| Share high education | 0.151 | 0.216 |
| Share of Covid-19 deaths | 0.00107 | 0.00217 |
|  | Älvkarleby | Gävle |
| Share foreign-born | 0.149 | 0.159 |
| Share high education | 0.155 | 0.216 |
| Share of Covid-19 deaths | 0.00630 | 0.00217 |
|  | Heby | Avesta, Gävle, Sala, Sandviken |
| Share foreign-born | 0.126 | 0.165 |
| Share high education | 0.145 | 0.178 |
| Share of Covid-19 deaths | 0.00161 | 0.00227 |

Table A9. Determinants of share of vaccinated 18-29-year-olds in treated and neighboring municipalities

|  | (1) | (2) | (3) | (4) |
|---|---|---|---|---|
| Treatment | 0.063** | 0.020 | 0.042 | 0.007 |
|  | (0.023) | (0.012) | (0.033) | (0.022) |
| Neighbor indicators | No | No | Yes | Yes |
| Share foreign-born | No | -0.552*** | No | -0.377 |
|  |  | (0.100) |  | (0.210) |
| Share high education | No | 0.485*** | No | 0.320 |
|  |  | (0.074) |  | (0.140) |
| COVID-19 deaths | No | -2.883 | No | -14.421 |
|  |  | (4.555) |  | (13.133) |
| Constant | 0.745*** | 0.768*** | 0.781*** | 0.834*** |
|  | (0.016) | (0.024) | (0.050) | (0.067) |
| $R^2$ | 0.344 | 0.897 | 0.734 | 0.974 |
| Moran's I (residuals) | -0.214 | -0.130 | -0.124 | -0.417** |
| Conley SE (treatment) | 0.029** | 0.009** | 0.016*** | 0.011 |

Notes: The dependent variable is the share of 18-29-year-olds vaccinated in week 49 in the 16 included municipalities. Ordinary least squares regressions controlling for Treatment (pre-booked appointments), Neighbor indicators (one dummy variable for each treated municipality, indicating its neighbors), as well as the control variables Share foreign-born, Share high education, and COVID-19 deaths. Moran's I for spatial residual autocorrelation and Conley standard errors accounting for spatial autocorrelation were computed assuming a maximum distance for spatial autocorrelation of 65 kilometers (the minimum distance from which all area centroids shared at least one neighbor) and a Bartlett kernel using the *acreg* package for Stata. * $p<0.1$, ** $p<0.05$, *** $p<0.01$.